\documentclass[aps,prd,twocolumn]{revtex4-1}

\usepackage{epsfig}

\begin{document}
\author{Yu Jiang$^{1}$, Hao Gong$^{2}$, Wei-min Sun$^{2,3}$ and Hong-shi Zong$^{2,3}$}
\address{$^{1}$ Center for Statistical and Theoretical Condensed Matter Physics,
Zhejiang Normal University, Jinhua City, Zhejiang
Province 321004, China}
\address{$^{2}$ Department of Physics, Nanjing University, Nanjing 210093, China}
\address{$^{3}$ Joint Center for Particle, Nuclear Physics and Cosmology, Nanjing 210093, China}
\title{The Wigner solution of quark gap equation at nonzero current quark mass and partial restoration of chiral symmetry at finite chemical potential}
\begin{abstract}
According to the generally accepted phase diagram of QCD, at low
temperature and high baryon number density the chiral phase
transition of QCD is of first order and the co-existence of the
Nambu-Goldstone phase and the Wigner phase should appear. This is in
conflict with the usual claim that the quark gap equation has no Wigner
solution in the case of nonzero current quark mass. In this paper
we analyze the reason why the Wigner solution does not exist in the
usual treatment and try to propose a new approach to discuss this
question. As a first step, we adopt a modified Nambu-Jona-Lasinio (NJL) model to study the Wigner solution at finite current quark mass. We then generalize this approach to the case of finite chemical potential and discuss partial restoration of chiral symmetry at finite chemical potential and compare our results with those in the normal NJL model.

\bigskip


\bigskip

\noindent E-mail: zonghs@chenwang.nju.edu.cn.

\bigskip

\noindent PACS Number(s): 12.38.Mh, 12.39.-x, 25.75.Nq
\end{abstract}
\maketitle

\section{Introduction}
Dynamical chiral symmetry breaking (DCSB) and confinement are two
fundamental features of Quantum Chromodynamics (QCD). It is
generally believed that with increasing temperature and baryon
number density strongly interacting matter will undergo a phase
transition from the hadronic matter to the quark-gluon plasma (QGP)
which is expected to appear in the ultrarelativistic heavy ion
collisions. These two phases are generally referred to as the
Nambu-Goldstone phase which is characterized by DCSB and confinement
of dressed quarks and the Wigner phase corresponding to QGP in which
chiral symmetry is partially restored and quarks are not confined.
Theoretically, these two phases are described by two different
solutions, the Nambu-Goldstone solution and the Wigner solution of
the quark gap equation. The existence of these two solutions
in the chiral limit (the current quark mass $m=0$) has been shown in
the framework of Dyson-Schwinger equation (DSE) approach of QCD
(see, for example, \cite{DSER1,DSER2}). However, it is a general
conclusion in the previous literature that when the current quark mass $m$ is
nonzero, the quark gap equation has only one solution which corresponds to
the Nambu-Goldstone phase and the solution corresponding to the
Wigner phase does not exist \cite{DSE1,DSE2}. This conclusion is hard to understand and one will naturally ask why the Wigner solution of the quark gap equation only exists in the chiral limit while does not exist at finite current quark mass.  
Furthermore, this conclusion is in fact not compatible with the current study of
chiral phase transition of QCD. In order to see this more clearly,
let us have a look at the generally accepted QCD phase diagram (see,
for example, Fig. 3 of Ref. \cite{QCD1}). As is shown in the QCD
phase diagram, it is generally believed that at low temperature and
high baryon number density the chiral phase transition of QCD is of
first order and the co-existence of the Nambu-Goldstone phase and
the Wigner phase should appear. It is well-known that in the real world
the current quark mass is nonzero. If one cannot find the Wigner
solution of the quark gap equation in the case of nonzero current quark mass, this will mean
that we cannot talk about the co-existence of these two phases. This
is obviously an unsettled and important problem in the study of QCD
phase transitions. The authors of Ref. \cite{zong1} first discussed
this problem and asked whether the quark gap equation has a Wigner solution
in the case of nonzero current quark mass. Subsequently, the authors
of Refs. \cite{DSE3,DSE4,DSE5} further investigated the problem of
possible multi-solutions of the quark gap equation. However, as far as we
know, this problem has not been solved satisfactorily in the
literature. In the present paper we try to propose a new approach to
investigate this problem.

The main motivation of the present work is to study the Wigner solution of the quark gap equation at finite current quark mass and provide a new viewpoint on partial restoration of chiral symmetry at finite chemical potential. 
This paper is organized as follows: in Sect. II we analyze the reason why in the previous literature the Wigner solution of the quark gap equation does not exist in the case of the finite current quark mass and propose a new approach to discuss this question. In Sect. III, based on such an approach,  we show in the framework of Nambu-Jona-Lasinio (NJL) model that the quark DSE has a Wigner solution at finite current quark mass. Then, in Sect IV we generalize this approach to the case of finite chemical potential to study partial restoration of chiral symmetry and compare our results with the corresponding ones in previous literature. The results are summarized in Sect. V. 

\section{Quark gap equation and its solutions}
\begin{figure}[h]
\includegraphics[width=8cm]{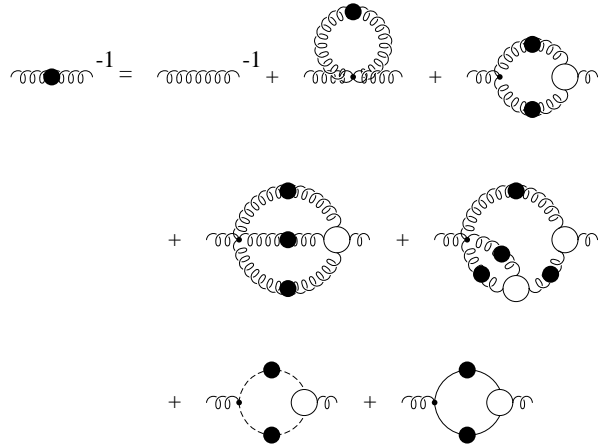}
\caption{The DSE for the dressed gluon propagator}\label{fig1}
\end{figure}

In order to illustrate our new approach more clearly, let us now
briefly recall the usual arguments which exclude the existence of
the Wigner solution of the quark gap equation when $m\neq0$. The quark DSE
under rainbow approximation reads as following (in the
present paper we will always work in Euclidean space, and take the number of flavours, $N_f=2$ and the number of colours, $N_c=3$)
\begin{equation}
G^{-1}(p)=G^{-1}_0(p)+\frac{4}{3}
\int\frac{d^4q}{(2\pi)^4}g^2D_{\mu\nu}(p-q)\gamma_\mu
G(q)\gamma_\nu,\nonumber\\\label{gap1}
\end{equation}
where $G(p)$ is the dressed quark propagator, $G_0(p)=(i\gamma\cdot
p+m)^{-1}$ is the free quark propagator, $g$ is the strong coupling
constant and $D_{\mu\nu}(q)$ is the effective dressed gluon propagator.
According to Lorentz structure analysis, one has
\begin{equation}
G^{-1}(p)=i{\not\!p}A(p^2)+B(p^2),\label{quark1}
\end{equation}
where $A(p^2)$ and $B(p^2)$ are scalar functions of $p^2$.
Substituting Eqs. (\ref{quark1}) into Eq.
(\ref{gap1}), one has
\begin{eqnarray}
[A(p^2)-1]p^2&=&\frac{4}{3}\int\frac{d^4q}{(2\pi)^4}
\frac{g^2D(p-q)A(q^2)}{q^2A^2(q^2)+B^2(q^2)}\nonumber\\
&&\times\left[p\cdot
q+2\frac{p\cdot(p-q)q\cdot(p-q)}{(p-q)^2}\right]\label{quark2},\\
B(p^2)&=&m+\int\frac{d^4q}{(2\pi)^4}
\frac{4g^2D(p-q)B(q^2)}{q^2A^2(q^2)+B^2(q^2)},\label{quark3}
\end{eqnarray}
where Landau gauge has been employed. From Eqs. (\ref{quark2}) and
(\ref{quark3}) one can find when $m=0$ there are two distinct
solutions for $B(p^2)$. One solution is $B(p^2)\neq0$ which
describes the Nambu phase, and the other one is $B(p^2)\equiv0$
which describes the Wigner phase. However, when $m\neq0$, it can be
easily seen that $B(p^2)\equiv0$ is not a solution of Eqs.
(\ref{quark2}) and (\ref{quark3}). From this observation one often
concludes that when $m\neq0$, the quark DSE has only one solution
corresponding to the Nambu phase and the Wigner solution does not
exist. Here, it should be noted that in obtaining this conclusion
one has assumed that the dressed gluon propagators in these two
phases are the same. However, since the features of these two phases
are so different, it is reasonable to expect that the behavior of
the dressed gluon propagator should be different in these two phases
(for example, in the familiar liquid-solid phase transition of
water, the effective interactions between molecules are different in
the two phases). To see this more clearly, let us look at the
graphical representation of the DSE for the dressed gluon propagator
given in Fig. \ref{fig1}.
From Fig. \ref{fig1} it can be seen that the quark propagator can
affect the gluon propagator through quark-loop insertions.
Therefore, in principle, since the quark propagators in Nambu phase
and Wigner phase are quite different, one naturally expects that the
gluon propagators in these two phases should be different, too.  Here we would like to stress that this observation is model independent. Besides, this observation has been verified in the study of quantum
electrodynamics in $2+1$ dimensions (QED$_3$) by using the coupled
DSE for the fermion and photon propagators with a range of
fermion-photon vertices \cite{DSE6} (QED$_3$ has many features
similar to QCD, such as spontaneous chiral symmetry breaking in the
massless fermion limit and confinement. Due to these reasons it can serve as a toy model of QCD). This indicates that one should
choose different forms of gluon propagator as input to solve the
quark propagators in the two different phases. Now, the key problem
is how to choose appropriate model gluon propagators as input to
calculate the dressed quark propagator in the Nambu phase and the
Wigner phase, respectively.

From Fig. \ref{fig1} it can be seen that one can formally split the full gluon propagator into two parts as following
\begin{eqnarray}
D_{\mu\nu}(q)&=&D^{\mathrm{YM}}_{\mu\nu}(q)+D^{Q}_{\mu\nu}(q),
\label{gluon2}
\end{eqnarray}
where $D^{\mathrm{YM}}_{\mu\nu}$ is the pure Yang-Mills part which
includes all diagrams without quark loop insertions (which is
usually called quenched gluon propagator in lattice QCD) and
$D^{Q}_{\mu\nu}$ is the quark-affected part which includes all
diagrams with quark loop insertions. Obviously, the pure Yang-Mills
part in Wigner phase should be same as that in Nambu phase, whereas
in principle the quark-affected parts in these two phases should be
different. At present it is impossible to calculate the two parts
$D^{\mathrm{YM}}_{\mu\nu}(q)$ and $D^{Q}_{\mu\nu}(q)$ from first
principle of QCD. So one has to resort to various nonperturbative
QCD models to express them phenomenologically.

Over the past few years, considerable progress has been made in the framework of the QCD sum rule \cite{Shifman}, which provides a successful description of various nonperturbative aspects of strong interaction physics. We naturally expect that it might provide some useful clue to the study of the model gluon propagator. From the QCD sum rule approach the lowest-order contribution of quark condensate to the gluon propagator is \cite{SumR1}
\begin{eqnarray}
\Delta_{\mu\nu}(p)&=&-g^2\int
d^4(y-z)\int\frac{d^4q}{(2\pi)^4}e^{i(p-q)\cdot(y-z)}\nonumber\\
&&\times\mbox{tr}\left[\gamma_\mu
\frac{-i{\not\!q}+m}{q^2+m^2}\gamma_\nu\langle
\bar\psi(y)\psi(z)\rangle\right]\nonumber\\
&\sim&-\delta_{\mu\nu}\frac{mg^2\langle\bar\psi\psi\rangle}{3p^2}+\cdots,
\label{gluon3}
\end{eqnarray}
where $\langle\bar\psi(y)\psi(z)\rangle$ is the non-local quark
condensate and $\langle\bar\psi\psi\rangle$ is the ordinary 
two-quark condensate, the ellipsis represents terms of higher orders in
$\frac{m^2}{p^2}$ which we neglect in the present work since we
limit our discussion to two light flavors $u$ and $d$. It is evident that the value of quark condensate $\langle {\bar \psi}\psi\rangle$ is different in the Nambu phase and Wigner phase. This makes the gluon propagators
in these two phases be different. Therefore, in the following
calculation we can phenomenologically identify $\Delta_{\mu\nu}(p)$
in Eq. (\ref{gluon3}) as a good approximation of the
$D^{Q}_{\mu\nu}(q)$ part in Eq. (\ref{gluon2}).

\section{NJL-like model and two distinct solutions at zero chemical potential}
Now, we should specify a model framework to calculate the quark
propagators in Nambu phase and the Wigner phase. The dressed quark propagators are the most elementary of the n-point Green functions of QCD. It is evident that the Dyson-Schwinger equations (DSEs) are the natural tool for investigating it in the continuum. In particular, it has been shown that DSEs are capable to describe the chiral phase transition and deconfinement phase transition at finite temperatures and chemical potential \cite{DSER2,Hou,Jiang,Fisher1,Fisher2,Qin}. However, as is shown in Ref. \cite{NJL1}, the Nambu-Jona-Lasinio (NJL)
model can capture the main physical features of QCD at finite temperature and chemical potential. For example, partial restoration of chiral symmetry, the critical end point and color superconductivity are all first studied in the framework of the NJL model. This is the reason why the NJL model is the most widely used QCD model in the study of QCD phase transition at finite temperature and chemical potential (although this model has two defects, namely, it can neither accommodate confinement nor is 
renormalizable). Therefore, as a first step, for simplicity in this paper we shall employ the NJL model to study the quark
propagators in Nambu phase and the Wigner phase. 

In the normal NJL model the following model gluon propagator
\begin{eqnarray}
g^2D_{\mu\nu}(p-q)&=&\delta_{\mu\nu}\frac{1}
{M_G^2}\theta(\Lambda^2-q^2),\label{gluon4}
\end{eqnarray}
is employed to calculate the quark propagator,
where $M_G$ is some effective gluon mass scale and $\Lambda$ serves
as a cutoff and is set to be $1.015$ GeV in Ref. \cite{NJL1}. This model gluon propagator concentrates on the infrared region of the interaction
which is believed to be vital for DCSB of QCD. With such a model gluon
propagator Eq. (\ref{gap1}) becomes
\begin{eqnarray}
&&i{\not\!p}A(p^2)+B(p^2)=i{\not\!p}+m+\frac{4}{3M_G^2}\nonumber\\
&&\times\int\frac{d^4q}{(2\pi)^4}\theta(\Lambda^2-q^2)
\frac{\gamma_\mu[-i{\not\!q}A(q^2)+B(q^2)]
\gamma_\mu}{A^2(q^2)q^2+B^2(q^2)}.\label{gap2}
\end{eqnarray}
The solution of Eq. (\ref{gap2}) is $A(p^2)\equiv1$ and
$B(p^2)\equiv M$ with $M$ being a constant satisfying the following
equation
\begin{eqnarray}
M&=&m+\frac{M}{3\pi^2M_G^2}D_1(M^2,\Lambda^2),\label{gap3}
\end{eqnarray}
where $D_1(M^2,\Lambda^2)=\Lambda^2-M^2\ln[1+\Lambda^2/M^2]$. From
Eq. (\ref{gap3}) it is easy to find that when $M_G^2<1/3\pi^2$, this
equation has two different solutions in the chiral limit. However,
when $m\neq0$, one could only find one solution, the Nambu solution,
which satisfies $M>0$. This result is consistent with the one derived 
from the analysis of the quark DSE under rainbow approximation. 

Obviously, the vacuum of the Wigner phase should be different from
that of the Nambu phase and their difference can be characterized by
the quark condensate which is regarded as the order parameter for
chiral phase transition. Therefore, the gluon propagator should be
different due to different values of quark condensate in these two
phases. In order to reflect this fact, we introduce the quark
condensate contribution (Eq. (\ref{gluon3})) to the gluon
self-energy and modify the effective gluon propagator in the normal
NJL model as following
\begin{eqnarray}
g^2D_{\mu\nu}(q)&&=\delta_{\mu\nu}\frac{1}
{M_G^2}\theta(\Lambda^2-q^2)-\delta_{\mu\nu}\frac{1}
{M_G^2}\frac{m\langle\bar\psi
\psi\rangle}{\Lambda_q^2}\nonumber\\
&&\times\frac{1}{M_G^2}\theta(\Lambda^2-q^2)
\label{gluon5}=\delta_{\mu\nu}\frac{1}
{M_\mathrm{eff}^2}\theta(\Lambda^2-q^2),\nonumber\\
\end{eqnarray}
where the first term in the right-hand-side of Eq. (\ref{gluon5}) is
the usual model gluon propagator employed in the NJL model which has
the same form in both Nambu phase and Wigner phase and can be
regarded as the pure Yang-Mills part $D^{YM}_{\mu\nu}(q)$ in the
present work; the second term, which is inspired by the result of
QCD sum rules \cite{SumR1}, is the leading order non-perturbative
contribution from quark condensate through quark loop insertions.
Here, it should be noted that according to the usual approximation
of NJL model, in obtaining the current quark mass dependent term of Eq. (\ref{gluon5}) we have taken all the momentum dependence of the effective interaction in momentum space as a constant. For this purpose, we have
introduced a momentum scale $\Lambda_q$ which reflects the large
distance behavior of QCD. For the external momentum squared much
larger than $\Lambda_q^2$ the current quark mass dependent term of 
Eq. (\ref{gluon5}) can be neglected, whereas for external momentum squared approaching $\Lambda_q^2$ the contribution of quark condensate which has been neglected in the normal NJL model must be considered. Just as will be shown below, the current mass dependent term in Eq. (10) plays an important role in searching for the Wigner solution at finite current quark mass and the study of partial restoration of chiral symmetry at finite chemical potential.   

\begin{figure}
\includegraphics[width=8cm]{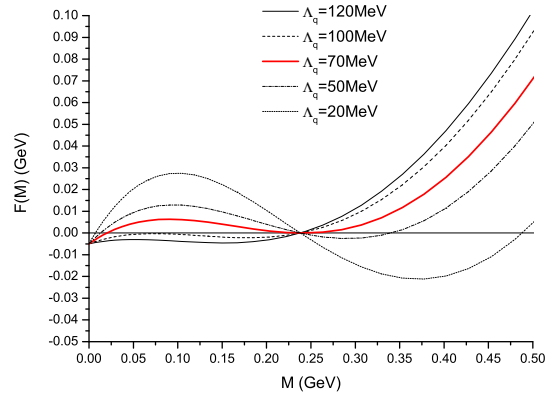}
\caption{Solutions of the gap equation with different $\Lambda_q$ ($m$ is fixed to be $5$ MeV)}\label{fig2}
\end{figure}

The value of $M_G$ which accounts for the pure Yang-Mills gauge
field contribution could be fixed by requiring the amount of the
intensity of the effective interaction to be
$M_\mathrm{eff}/\Lambda=0.17$ for the Nambu solution $M=238$ MeV
which is determined by fitting the observables such as pion decay
constant and pion mass (in the present paper we set the current
quark mass $m=5$ MeV) \cite{NJL1}.
With the modified gluon propagator given by Eq. (\ref{gluon5}), the
quark gap equation Eq. (\ref{gap3}) becomes
\begin{eqnarray}
M&=&m+\frac{M}{3\pi^2}\left[\frac{1}{M^2_G}+\frac{1}
{M^2_G}\frac{3MmD_1(M^2,\Lambda^2)}
{2\pi^2\Lambda_q^2}\frac{1}{M^2_G}\right]\nonumber\\
&&\times D_1(M^2,\Lambda^2).\label{gap4}
\end{eqnarray}
Now let us turn to the the calculation of  Eq. (\ref{gap4}). To
illustrate how the solution of Eq. (\ref{gap4}) varies with
different $\Lambda_q$, let us define
\begin{eqnarray}
F(M)&=&M-m-\frac{M}{3\pi^2}\left[\frac{1}{M^2_G}+\right.\nonumber\\
&&\left.\frac{1} {M^2_G}\frac{3MmD_1(M^2,\Lambda^2)}
{2\pi^2\Lambda_q^2}\frac{1}{M^2_G}\right]C(M^2,\Lambda^2),\nonumber\\
\end{eqnarray}
and the solution of $F(M)=0$ is just the solution of the quark gap
equation (Eq. (\ref{gap4})). In Fig. \ref{fig2} $F(M)$ is plotted as a
function of $M$ with different $\Lambda_q$. From Fig. \ref{fig2} it
can be seen that when $\Lambda_q$ is larger than about $100$ MeV,
the equation $F(M)=0$ has only one solution $M=238$ MeV, which is
similar to the situation discussed in Ref. \cite{DSE3}. When
$\Lambda_q<100$ MeV, the equation $F(M)=0$ has three solutions.
Specifically, when $70 \mathrm{MeV}<\Lambda_q<100$ MeV, one solution
is the required Nambu solution $M=238$ MeV, and the other two
solutions are all smaller than it; when $\Lambda_q<70 \mathrm{MeV}$,
among the two solutions other than the Nambu one, one is smaller
than it and the other one is larger than it. Here we note that
physical observables require the Nambu solution to be $M=238$ MeV
and the stability condition of the Nambu solution would exclude the
existence of solutions larger than it. Therefore, the parameter
$\Lambda_q$ should be constrained within the range $70
\mathrm{MeV}\leq\Lambda_q\leq100 \mathrm{MeV}$. For $\Lambda_q$ in
this range, the smallest solution of Eq. (\ref{gap4}) is not very
large compared with the current quark mass, and when the current quark
mass $m$ tends to zero, this solution will continuously tend to
zero, which is just the Wigner solution in the chiral limit.
Therefore, this solution might be identified as the Wigner solution
in the case of $m\neq 0$ which describes the perturbative dressing
effect.

\begin{figure}
\includegraphics[width=8cm]{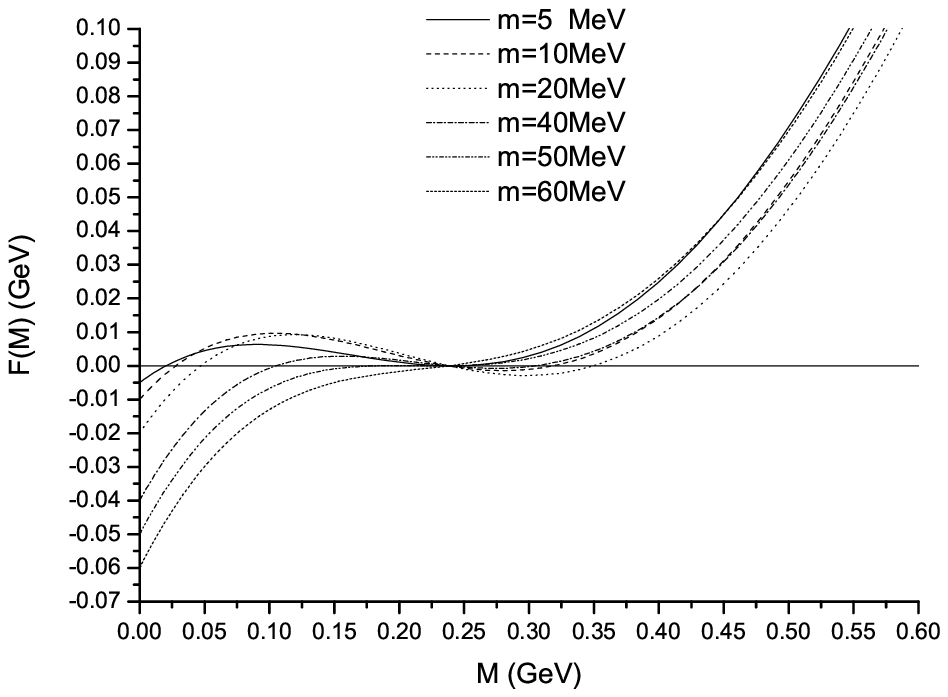}
\caption{Solutions of the gap equation with different $m$ ($\Lambda_q$ is fixed to be $70$ MeV)}\label{fig3}
\end{figure}

The result in Fig. \ref{fig2} shows that the scale at which the
current quark mass dependent term of Eq. (\ref{gluon5}) would affect the effective interaction can change
the pattern of the solutions of the quark gap equation. If the current quark mass dependent term plays an important role in the infrared region in the
effective interaction ($\Lambda_q<100$ MeV), then the intensity of
the pure Yang-Mills field would be weakened and the Wigner solution
will appear. On the contrary, when $\Lambda_q>100$ MeV, the pure
Yang-Mills part would be dominating and strong in the infrared
region, and therefore the Wigner solution cannot exist due to strong
interaction. From physical consideration we choose $\Lambda_q=70$
MeV, because in this case the gap equation has just two solutions
which can be identified as the Nambu solution and the Wigner
solution. For $\Lambda_q=70$ MeV, we plot the $F(M)$ versus $M$
curve for different current quark mass $m$ in Fig. \ref{fig3}. It
can be seen that as $m$ increases, the effective mass $M$ of dressed quark in Wigner phase will increase and at last coincide with the Nambu solution when $m\sim60$ MeV.

\begin{center}
Table I The Nambu and Wigner solution
\begin{tabular*}{8cm}{l@{\extracolsep{\fill}}*{3}{c}}\hline\hline
$\Lambda_q=70$ (MeV) &~ M (GeV)&$-\langle\bar\psi\psi\rangle
(\mbox{GeV}^3)$&$\mathcal{P} (\mbox{GeV}^4)$\\
\hline Nambu phase&$0.238$&$3.13\times10^{-2}$&$1.797\times10^{-3}$\\
\hline Wigner phase&$0.02$&$3.16\times10^{-3}$&$3.06\times10^{-6}$\\
\hline\hline
\end{tabular*}
\end{center}

As usual, the quark condensate is defined as
\begin{eqnarray}
\langle\bar\psi\psi\rangle&=&-\int\limits^{\Lambda}
\frac{d^4p}{(2\pi)^4}\mbox{Tr}[G(p)]=-\frac{3MD_1(M^2,\Lambda^2)}{2\pi^2}
\end{eqnarray}
and its value for the two solutions is listed in Table I (In Table I we list the solution of Eq. (\ref{gap4}) with $\Lambda_q=70$ MeV and $m=5$ MeV)
. It can be seen that the value of the quark condensate in Nambu phase is
larger than that of Wigner phase by one order of magnitude, which
represents DCSB of Nambu phase. Here it should be pointed out that
the quark condensate of Wigner solution is small but non-zero
because it reflects the explicit chiral symmetry breaking due to
non-zero current quark mass.

Of course, in order to determine which solution is the real one, one should
compare the pressure (thermodynamical potential) of the different
solutions. The vacuum pressure $\mathcal{P}$ of the two solutions
are also listed in Table I which is calculated via ``steepest
descent'' approximation as following \cite{DSE2}
\begin{equation}
\mathcal{P}=\int\limits^\Lambda\frac{d^4p}
{(2\pi)^4}\mbox{Tr}\Bigg\{\ln [G^{-1}(p)G_0(p)]+\frac{1}{2}
\left[G_0^{-1}(p)G(p)-1\right]\Bigg\}.\label{pressure1}
\end{equation}
From Table I it can be seen that the vacuum pressure of Nambu phase
is much larger than that of the Wigner phase (more than two orders
of magnitude), which means Nambu phase is more stable than the
Wigner phase when temperature and density are zero. The vacuum
pressure difference of the two phases can be regarded as the bag
constant $B_\mathrm{bag}$ and the results in Table I correspond to
$B_\mathrm{bag}\sim(206\mathrm{MeV})^4$ which is consistent with the
value used in the literature \cite{DSE1}. One may expect that with increasing
temperature and/or density this quantity may change and chiral phase
transition would happen. We will discuss this question in the next section.

\section{partial restoration of chiral symmetry at finite chemical potential}
Now we can generalize the previous treatment to the case of finite density. The quark propagator at finite quark chemical potential $\mu$ could be expressed as following 
\begin{eqnarray}
G^{-1}(p,\mu)&=&iA{\not\!p}+B-C\mu\gamma_4,
\end{eqnarray}
where $A(\vec p^2,p_4,\mu)$, $B(\vec p^2,p_4,\mu)$ and $C(\vec
p^2,p_4,\mu)$ are scalar functions of $\vec p^2$, $p_4$ and $\mu$.
With the model gluon propagator in Eq. (\ref{gluon5}) the DSE of
quark propagator at finite chemical potential is
\begin{eqnarray}
&&iA{\not\!p}+B-C\mu\gamma_4=i{\not\!p}+m-\mu\gamma_4\nonumber\\
&&+\frac{4}{3M^2_{\mathrm{eff}}}\int\frac{d^4q}{(2\pi)^4}\frac{2iA{\not\!q}+4B-2C\mu\gamma_4}{A^2q^2+B^2-C^2\mu^2+2iAC\mu q_4}.
\end{eqnarray}
From the above equation one could easily find the solution should be
$A=1$ and $B$ and $C$ are constant. The constant $C\mu$ plays the
role of effective chemical potential and therefore let us set
$\mu^*=C\mu$ and $B=M$ which satisfy the following combined
equations
\begin{eqnarray}
M&=&m+\frac{4}{3M^2_{\mathrm{eff}}}\int\frac{d^4q}{(2\pi)^4}
\frac{4M}{q^2+M^2-\mu^{*2}+2i\mu^*
q_4}\nonumber\\
&=&m+\frac{4M}{3M^2_{\mathrm{eff}}\pi^3}\int\limits_0^\Lambda d|\vec
q|\frac{\vec q^2}{E_{qM}}
\left[\arctan\left(\frac{\sqrt{\Lambda^2-\vec
q^2}}{E_{qM}+\mu^*}\right)\right.\nonumber\\
&&\left.+\arctan\left(\frac{\sqrt{\Lambda^2-\vec
q^2}}{E_{qM}-\mu^*}\right)\right],\label{quark4}\\
\mu^*&=&\mu-\frac{2\rho(\mu^*)}{3N_cN_fM^2_{\mathrm{eff}}},\label{quark5}
\end{eqnarray}
with quark number density $\rho(\mu^*)$ defined as following \cite{zong2}
\begin{eqnarray}
\rho(\mu^*)&=&-N_cN_f\int\frac{d^4q}{(2\pi)^4}
\mbox{tr}\big[G(q,\mu)\gamma_4\big],
\end{eqnarray}
and $E_{qM}=\sqrt{\vec q^2+M^2}$.
\begin{figure}[t]
\includegraphics[width=8cm]{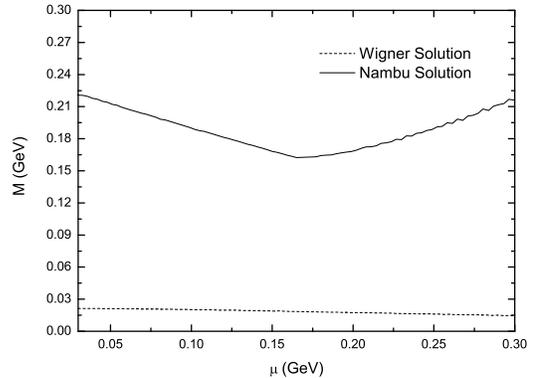}
\caption{Solutions of the gap equation at finite chemical potential}\label{fig4}
\end{figure}
The Eqs. (\ref{quark4}) and (\ref{quark5}) are numerically solved
and the results are shown in Fig. \ref{fig4}. From Fig. \ref{fig4} One could find the effective mass of the dressed quark in the Wigner phase decreases with increasing $\mu$, which means with
increasing density the dressing effect of quarks becomes more and
more weak. On the other hand, the corresponding one in Nambu phase decreases with increasing $\mu$ until $\mu\sim 160$ MeV, and when 
$\mu>160$ MeV the effective mass of the dressed quark in Nambu phase increases with increasing $\mu$. 

\begin{figure}[t]
\includegraphics[width=8cm]{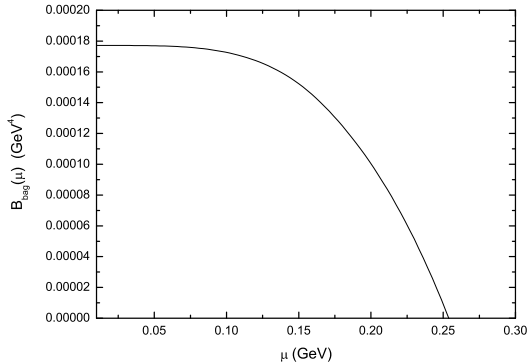}
\caption{The $B_{\rm{bag}}(\mu)$ at finite $\mu$}\label{fig5}
\end{figure}

In the previous literature (see, e.g., Refs. \cite{Fisher2,He}) one usually employ the maximum of the susceptibility $\frac{\partial{\langle\bar{\psi}\psi\rangle}}{\partial m}$ to determine the transition temperature. In fact, a more reliable criterion for the chiral phase transition is the  pressure difference of the Nambu phase and the Wigner phase, i.e., the bag
constant $B_{\rm{bag}}(\mu)$. The pressure density of the two solution at finite chemical potential could be calculated as following \cite{zong2}
\begin{eqnarray}
\mathcal{P}(\mu)&=&\mathcal{P}(\mu=0)+\int_0^\mu
d\mu^\prime\rho(\mu^\prime),
\end{eqnarray}
where the pressure density of the vacuum $\mathcal{P}(\mu=0)$ can be
calculated through Eq. (\ref{pressure1}). The $B_{\rm{bag}}(\mu)$ is
plotted in Fig. \ref{fig5} in which one can see when $\mu<\mu_c=260$ MeV the
Nambu solution is more stable and when $\mu>\mu_c=260$ MeV the Wigner
solution is more stable. At $\mu_c=260$ MeV the pressure of the two
phases is equal and the two phases could co-exist at this point. Here it should be noted that no one has calculated the $B_{\rm{bag}}(\mu)$ in the case of nonzero current quark mass in the past. This is due to lack of knowledge about the Wigner solution of the quark gap equation at finite current quark mass in the previous literature.

Here it is interesting to compare our results with those of the normal NJL model. The first-order phase transition point $\mu_c$ in our modified NJL model is smaller than the one obtained in the normal NJL model which is about 354 MeV or 500 MeV corresponding to different parameters \cite{NJL2}. It should also be pointed out that in the normal NJL model, the second solution appears
when $\mu$ is big enough \cite{NJL1}, but the magnitude of this
solution at the phase transition point is much bigger (about 110 MeV
or 130 MeV, see Ref. \cite{NJL2}) than the Wigner solution obtained
in the present paper (about 15 MeV). In addition, we want to stress that the
Wigner solution at finite current quark mass in the normal NJL model is due to density effect. When the chemical potential tends to zero, this solution
disappears. This shows $\mu=0$ is a singularity of the Wigner solution 
at finite current quark mass. If this is real, it means that one cannot study the
Wigner solution by means of small $\mu$ expansion, while the method of small $\mu$ expansion is a usually employed one in the study of lattice QCD at finite density.

\section{Summary}

To summarize, based on the general analysis that the dressed gluon
propagator in Wigner phase should be different from that in Nambu
phase, we introduce the contribution of quark condensate to the
gluon propagator and investigate the solution of quark DSE in the
case of nonzero current quark mass. With such a modified model gluon
propagator, in the framework of NJL model we show that the quark
DSE indeed has a Wigner solution in the case of nonzero current quark mass.
We then generalize this approach to the case of finite chemical potential and discuss partial restoration of chiral symmetry at finite chemical potential. From the calculated result of the bag constant we find that  
when $\mu<\mu_c=260$ MeV the Nambu solution is more stable and when $\mu>\mu_c=260$ MeV the Wigner solution is more stable. At $\mu_c=260$ MeV the pressure of the two phases is equal and the two phases could co-exist at this point. We also compare our results with those of the normal NJL model.
It is found that the first-order phase transition point $\mu_c$ in our modified NJL model is smaller than the one obtained in the normal NJL model which is about 354 MeV or 500 MeV corresponding to different parameters. 
In addition, in the normal NJL model, the second solution appears
when $\mu$ is big enough, but the magnitude of this
solution at the phase transition point is much bigger (about 110 MeV
or 130 MeV) than the Wigner solution obtained
in the present paper (about 15 MeV). 
Finally we would like to point out that the results obtained in this paper 
are based on a simple NJL model. It is well-known that the NJL model is
far from QCD. In order to obtain a more solid result, we need to further discuss this problem in the framework of a model with better QCD foundation,
such as DSE of QCD \cite{DSER2}.

\acknowledgments

We thank A. Bashir for helpful discussion. This work is supported in part by the Postdoctoral Science
Foundation of China (under Grant No. 20100471308), the National
Natural Science Foundation of China (under Grant Nos 11105122,
10935001 and 11075075) and and a project funded by the Priority Academic Program Development of Jiangsu Higher Education Institution.

\end{document}